\documentstyle[aps,times,epsf,multicol]{revtex}

\renewcommand{\narrowtext}
{\begin{multicols}{2} \global\columnwidth20.5pc}

\begin{document}

\draft

\title{Self-Duality in Superconductor-Insulator Quantum Phase
Transitions}
\author{Adriaan M. J. Schakel} 
\address{Low Temperature Laboratory, Helsinki University of Technology,
P.O. Box 2200, FIN-02015 HUT, Finland \\ and \\ Institut f\"ur
Theoretische Physik, Freie Universit\"at Berlin, Arnimallee 14, 14195
Berlin, Germany }

\date{\today}

\maketitle

\begin{abstract}
It is argued that close to a Coulomb interacting quantum critical point,
the interaction between two vortices in a disordered superconducting
thin film separated by a distance $r$ changes from logarithmic in the
mean-field region to $1/r$ in the region dominated by quantum critical
fluctuations.  This gives support to the charge-vortex duality picture
of the observed reflection symmetry in the current-voltage
characteristics on both sides of the transition.
\end{abstract}


\narrowtext

One of the most intriguing results found in experiments on quantum phase
transitions in superconducting films, 2-dimensional Josephson-junction
arrays \cite{Delft}, quantum Hall systems \cite{IVQH}, and 2-dimensional
electron systems \cite{KSSMF} is the striking similarity in the
current-voltage ($I$-$V$) characteristics on both sides of the
transition.  By interchanging the $I$ and $V$ axes in one phase, an
$I$-$V$ characteristic of that phase at a given value of the applied
magnetic field (in superconducting films, 2-dimensional
Josephson-junction arrays, and quantum Hall systems) or charge-carrier
density (in 2-dimensional electron systems) can be mapped onto an
$I$-$V$ characteristic of the other phase at a different value of the
magnetic field or charge-carrier density.  This reflection symmetry
hints at a deep connection between the conduction mechanisms in the two
phases that can be understood by invoking a duality transformation
\cite{MPAFisher,WeZe}.  Whereas the conducting phase is most succinctly
described in terms of charge carriers of the system, the insulating
phase is best formulated in terms of vortices, which behave as quantum
point particles in these systems.  The duality transformation links the
two surprisingly similar looking descriptions.

There appears to be, however, one disturbing difference.  Whereas 
charges interact via the usual 3-dimensional $1/r$ Coulomb potential,
vortices are believed to interact via a logarithmic potential--at least
for distances smaller than the transverse magnetic penetration depth
$\lambda_\perp$, which is typically larger than the sample size
\cite{BMO}.  This is disturbing because the difference should spoil the
experimentally observed reflection symmetry.

It is this fundamental problem we wish to address in this Letter.  It
will be shown that close to a Coulomb interacting quantum critical point
(CQCP), the interaction between vortices in disordered superconducting
films changes from logarithmic in the mean-field region to $1/r$ in the
region dominated by quantum critical fluctuations.  This conclusion is
an exact result, depending only on the presence of a CQCP.

A common characteristic of the systems mentioned is, apart from
impurities, the presence of charge carriers confined to move in a
2-dimensional plane.  As the $1/r$ Coulomb repulsion between 
charges is genuine 3-dimensional, we assume this interaction not to be
affected by what happens in the film, which constitutes a mere slice of
3-dimensional space.  In contrast to this, the interaction between
vortices is susceptible to the presence of a CQCP.  This is because the
vortex interaction is a result of currents around the vortex cores which
are confined to the plane.

As starting point, we take the observation (for a review, see Ref.\
\cite{SGCS}) that close to a CQCP, the electric field $E$ scales with
the correlation length $\xi$ as $E \sim \xi_t^{-1} \xi^{-1} \sim
\xi^{-(z+1)}$.  Here, $\xi_t$ denotes the correlation time, indicating
the time period over which the system fluctuates coherently, and $z$ is
the dynamic exponent.  Thus conductivity measurements \cite{YaKa,KSSMF}
close to a CQCP collapse onto a single curve when plotted as function of
the dimensionless combination $\delta^{\nu (z+1)}/E$, where
$\delta=(K-K_{\rm c})/K_{\rm c}$ measures the distance from the critical
point $K_{\rm c}$, and $\nu$ is the correlation length exponent, $\xi
\sim \delta^{- \nu}$.  (For a field-controlled transition, $K$ stands
for the applied magnetic field, while for a density-controlled
transition it stands for the charge-carrier density.)  The scaling of
the electric field with the correlation length expresses the more
fundamental result that the anomalous scaling dimension $d_{\bf A}$ of
the magnetic vector potential ${\bf A}$ is unity, $d_{\bf A} = 1$.

In addition, because the magnetic vector potential always appears in the
gauge-invariant combination $\nabla - q {\bf A}$, the anomalous scaling
dimension of the electric charge $q$ of the charge carriers times the
vector potential is unity too, $d_{q {\bf A}} = 1$.  Writing the
anomalous scaling dimension of the vector potential as a sum $d_{\bf A}
= d^0_{\bf A} + \case{1}{2} \eta_{\bf A}$ of its canonical scaling
dimension $d^0_{\bf A} = \case{1}{2} (d + z -2)$, obtained by simple
power counting, and (half) the critical exponent $\eta_{\bf A}$,
describing how the correlation function decays at the critical point, we
conclude that $d_q = d^0_q - \frac{1}{2}\eta_{\bf A}$.  Here, $d^0_q = 1
- d^0_{\bf A}$ stands for the canonical scaling dimension of the
electric charge.  Now, for a $1/r$ Coulomb potential, the charge scales
as $q^2 \sim \xi^{1-z}$ independent of the number $d$ of space
dimensions \cite{FGG}.  Combined with the previous result, this fixes
the value of the critical point decay exponent $\eta_{\bf A}$ in terms
of the number of space dimensions and the dynamic exponent:
\begin{equation} \label{eta}
\eta_{\bf A}=5 -d - 2z.
\end{equation} 

In Ref.\ \cite{FGG} it was further argued that in the presence of
disorder, the electric charge is finite at a CQCP, so that $z=1$.  This
prediction was first confirmed for disordered superconducting films
\cite{HPsu1}, and subsequently also for 2-dimensional Josephson-junction
arrays \cite{Delft}, quantum Hall systems \cite{WET}, and 2-dimensional
electron systems \cite{KSSMF}.  With $z=1$, the value of the
critical-point decay exponent becomes $\eta_{\bf A}=1$ in $d=2$.  As we
will now demonstrate, this leads to a qualitative change in the
interaction potential between two vortices from logarithmic in the
mean-field region, where $\eta_{\bf A} = 0$, to $1/r$ in the vicinity of
the CQCP, where $\eta_{\bf A}=1$.

To set the stage, let us first consider a bulk superconductor with two
static vortices directed along the $x_3$-axis and separated a distance
$r$.  For our purposes, the effective phase-only \cite{NgSu} Hamiltonian
${\cal H}_{\rm eff} = (\rho_{\rm s}/2 m^2) (\nabla \varphi - q {\bf
A})^2$ in terms of the phase $\varphi$ of the superconducting order
parameter---the so-called Anderson-Bogoliubov mode---suffices (for
reviews, see Ref.\ \cite{reviews}).  Here, $\rho_{\rm s}$ is the
superconducting mass density, which scales as $\rho_{\rm s} \sim
\xi^{2-(d+z)}$ \cite{FF}, and $m$ is the mass of the charge carriers.
The interaction potential can be extracted from the magnetic part of the
effective action $S_{\rm mag}$.  Written as a functional integral over
the the magnetic vector potential, it is given in the Coulomb gauge
$\nabla \cdot {\bf A}=0$ by
\begin{mathletters}
\begin{equation} \label{mag}
{\rm e}^{i S_{\rm mag}} = \int \mbox{D} {\bf A} \, {\rm e}^{i \int
\mbox{d} t \, \mbox{d}^3 x [-\frac{1}{2}(\nabla \times {\bf A} - {\bf
B}^{\rm P})^2 - \frac{1}{2} \lambda^{-2}{\bf A}^2 ]},
\end{equation} 
with $\lambda$ the magnetic penetration depth, which is related to
$\rho_{\rm s}$ via $\lambda^{-2} = q^2\rho_{\rm s}/m^2$.  The mass
term is generated through the Anderson-Higgs mechanism by integrating
out the phase mode $\varphi$.  The so-called plastic field ${\bf B}^{\rm
P}$ \cite{GFCM}
\begin{equation}  \label{BP}
B_i^{\rm P} = - \Phi_0 \sum_{\alpha} \int_{C_\alpha} \mbox{d} x_i^\alpha
\, \delta({\bf x} - {\bf x}^\alpha),
\end{equation} 
\end{mathletters}
with $\Phi_0 = 2 \pi/q$ the magnetic flux quantum in units where the
speed of light and Planck's constant $\hbar$ is set to unity, describes
the two vortices located along the lines $C_\alpha$ ($\alpha=1,2$).

Note that since the anomalous scaling dimension of the magnetic vector
potential is unity, the dimension of the Maxwell term is 4, implying
that in $d=2$ it is an irrelevant operator in the renormalization-group
sense.  This term is, however, important when considering the
interaction between vortices.

To facilitate the calculation in the case of a superconducting film
below, we linearize the first term in Eq.\ (\ref{mag}) by introducing an
auxiliary field $\tilde{\bf h}$ via a Hubbard-Stratonovich
transformation to obtain the combination $i (\nabla \times {\bf A} -
{\bf B}^{\rm P}) \cdot \tilde{\bf h} - \case{1}{2} \tilde{\bf h}^2$.
After integrating out the magnetic vector potential, we arrive at a form
appropriate for a dual description in terms of magnetic vortices rather
than electric charges \cite{KKS}
\begin{equation} \label{3d}
{\rm e}^{i S_{\rm mag}} = \int \mbox{D} \tilde{\bf h} \, {\rm e}^{i \int
\mbox{d} t \, \mbox{d}^3 x [-\case{1}{2} \lambda^2 (\nabla \times
\tilde{\bf h})^2 - \case{1}{2}\tilde{\bf h}^2 - i \tilde{\bf h} \cdot
{\bf B}^{\rm P} ] }.
\end{equation} 
Physically, $\tilde{\bf h}$ represents ($i$ times) the fluctuating local
induction; it satisfies the condition $\nabla \cdot \tilde{\bf h}=0$.
The vortices couple with a coupling constant $g=\Phi_0/\lambda$
independent of the electric charge to $\tilde{\bf h}$.  Observe the
close similarity between the original (\ref{mag}) and the dual form
(\ref{3d}).  This becomes even more so when an external electric current
${\bf j}^{\rm P}$ is coupled to the ${\bf A}$ field by including a term
$- {\bf A} \cdot {\bf j}^{\rm P}$ in Eq.\ (\ref{mag}), and ${\bf
B}^{\rm P}$ describing the vortices is set to zero there.  

Integrating out the local induction, one obtains the well-known
Biot-Savart law for the interaction potential $S_{\rm mag} = - \int
\mbox{d} t V$ between two static vortices in a bulk superconductor
\cite{deGennes},
\begin{eqnarray} \label{V3d}
V(r) &=& \frac{1}{2\lambda^2} \int \mbox{d}^3 x \, \mbox{d}^3 y B_i^{\rm
P}({\bf x}) G({\bf x} - {\bf y}) B_i^{\rm P}({\bf y}) \nonumber \\ &=&
\frac{g^2}{4 \pi} \int_{C_1} \int_{C_2} \mbox{d} {\bf l}^1 \cdot
\mbox{d} {\bf l}^2 \; \frac{{\rm e}^{-R/\lambda}}{R} \nonumber \\ &=& -
\frac{g^2}{2 \pi} L \left[\ln(r/2\lambda) + \gamma \right] + {\cal
O}(r/\lambda)^2
\end{eqnarray} 
where we ignored the self-interaction.  In Eq.\ (\ref{V3d}), $G({\bf
x})$ is the correlation function whose Fourier transform reads $G({\bf
k})=1/({\bf k}^2 + \lambda^{-2} )$, $R$ denotes the distance between the
differential lengths $\mbox{d} {\bf l}^1$ and $\mbox{d} {\bf l}^2$, $L$
is the length of each of the two vortices, and $\gamma$ is Euler's
constant.  For distances smaller than the magnetic penetration depth,
which is the length scale for variations in the current and the magnetic
field, the interaction is logarithmic as in a superfluid.  If the system
size is smaller than $\lambda$, it will replace $\lambda$ as infra-red
cutoff in the logarithm, and there will be no reference to the electric
charge anymore.

To describe magnetic vortices in a film of thickness $w$ \cite{Pearl},
the bulk result (\ref{3d}) has to be adjusted in two ways
to account for the fact that both the vortices and the screening
currents, which produce the second term in (\ref{3d}), are confined to
the plane.  This is achieved by including a Dirac delta function $w
\delta(x_3)$ in the second and third term.  Instead of Eq.\ (\ref{3d}),
one then arrives at the interaction potential \cite{Pearl,deGennes}
\begin{mathletters}
\begin{eqnarray} \label{V2d}
V_\perp(r) &=& \frac{1}{2 \lambda_\perp} \int \mbox{d}^2 x_\perp
\mbox{d}^2 y_\perp B_\perp^{\rm P}({\bf x}_\perp) G_\perp({\bf x}_\perp
- {\bf y}_\perp) B_\perp^{\rm P}({\bf y}_\perp) \nonumber \\ &=&
-\frac{g_\perp^2}{2 \pi} \left[ \ln(r/4 \lambda_\perp) + \gamma\right] +
{\cal O}(r/\lambda_\perp)^2,
\end{eqnarray} 
where $B_\perp^{\rm P} = - \Phi_0 \sum_{\alpha} \delta({\bf x}_\perp -
{\bf x}_\perp^\alpha)$ describes the vortices in the film with
coordinates ${\bf x}_\perp$, $\lambda_\perp = \lambda^2/w$ is the
transverse magnetic penetration depth, $g_\perp^2 =
\Phi_0^2/\lambda_\perp$ the coupling constant squared, and
\begin{eqnarray} 
G_\perp({\bf x}_\perp) &=& \int \mbox{d} x_3 \, G_\perp({\bf x}_\perp,
x_3) \nonumber \\ &=& \int \frac{\mbox{d}^2 k_\perp}{(2 \pi)^2} \, {\rm
e}^{- i {\bf k}_\perp \cdot {\bf x}_\perp} G_\perp({\bf k}_\perp,0),
\end{eqnarray} 
\end{mathletters}
with $G_\perp({\bf k}_\perp,0) = 2/ k_\perp (2 k_\perp
+\lambda_\perp^{-1})$.  For small distances, the interaction is seen to
be identical to that in a bulk superconductor \cite{Pearl}, and also to
that in a superfluid film.  As in the bulk, the vortex coupling constant
$g_\perp$ in the film is independent of the electric charge, $g_\perp^2
= \Phi_0^2/\lambda_\perp = (2 \pi)^2 \rho_{\rm s} w /m^2$, with
$\rho_{\rm s}$ the bulk superconducting mass density.

The above results are valid in the mean-field region, where $\eta_{\bf
A}=0$.  In the critical region governed by a CQCP, the value of this
exponent is unity, and the correlation function becomes
\begin{equation} \label{Gr} 
G_\perp({\bf k}_\perp,0) = \frac{2}{k_\perp } \frac{Z_{\bf
A}}{ 2 k_\perp +\lambda_\perp^{-1}},
\end{equation} 
with $Z_{\bf A} \sim k_\perp^{\eta_{\bf A}}$ the field renormalization
factor.  Because the magnetic vector potential and the local induction
renormalize in the same way, their renormalization factor is identical.
Due to this extra factor, the interaction between two vortices in the
film takes the form of a $1/r$ Coulomb potential
\begin{equation}  \label{renV}
V_\perp(r) = \frac{g_\perp^2}{2 \pi} \frac{a}{r},
\end{equation} 
where $a$ is some microscopic length scale which accompanies the
renormalization factor $Z_{\bf A}$ for dimensional reasons
\cite{Goldenfeld}.  

Since the electric charge is finite at the CQCP, the penetration depth
$\lambda_\perp \propto 1/\rho_{\rm s}$ scales with the correlation
length as $\lambda_\perp \sim \xi$.  In the correlation function
(\ref{Gr}) we thus have the combination $1/(2 k_\perp +\xi^{-1})$ which
should be compared with $1/({\bf k}^2 + \xi^{-2})$ for a bulk
superconductor.

The absence of any reference to the electric charge in the renormalized
and bare interaction (at least for small enough systems) implies that
the same results should be derivable from our starting Hamiltonian
restricted to two dimensions and with $q$ set to zero: ${\cal H}_\perp =
(\rho_{\rm s} w/2 m^2) (\nabla_\perp \varphi - \bbox{\varphi}_\perp^{\rm
P})^2$.  The plastic field $\bbox{\varphi}_\perp^{\rm P}$, with
$\nabla_\perp \times \bbox{\varphi}_\perp^{\rm P} = -2 \pi \sum_\alpha
\delta({\bf x}_\perp - {\bf x}_\perp^\alpha)$ describes vortices in a
superfluid \cite{GFCM}.  It is obtained from the description involving
the plastic field $B_\perp^{\rm P}$ by a canonical transformation of the
vector potential.  By directly integrating out the Anderson-Bogoliubov
mode, and ignoring the $k_\perp$ dependence of $\rho_{\rm s}$, which is
valid outside of the critical region, one easily reproduces the bare
interaction potential (\ref{V2d}).  The renormalized interaction
(\ref{renV}) is obtained by realizing that the anomalous scaling
dimension of the superconducting mass density is $d_{\rho_{\rm s}} =
(d+z)-2$ \cite{FF}, so that in our case $\rho_{\rm s} \sim k_\perp$.  In
other words, the extra factor of $k_\perp$ that came in via the
renormalization factor $Z_{\bf A}$ in our first calculation to produce
the $1/r$ potential, comes in via $\rho_{\rm s}$ here \cite{note}.

A similar change in the $r$-dependence of the interaction between two
vortices upon entering a critical region has been observed numerically
in the 3-dimensional Ginzburg-Landau model \cite{OlTe}.  Near the
charged fixed point of that theory, $\eta_{\bf a}=1$ \cite{HeTe}, as in
our case.  

This is a very pleasing coincidence as the (2+1)-dimensional
Ginzburg-Landau model constitutes the dual formulation of the system.
To appreciate the basic elements of the dual theory, note that the {\it
dynamics} of the charged degrees of freedom is described by the
effective Lagrangian
\begin{equation} \label{effprime}
{\cal L}_{\perp,{\rm eff}} = \frac{\rho_{\rm s}w}{2m^2}\left[
\frac{1}{c^2} (\partial_t\varphi + \varphi_t^{\rm P})^2 -(\nabla_\perp
\varphi - \bbox{\varphi}_\perp^{\rm P})^2 \right],
\end{equation} 
with $c$ the speed of sound.  In accord with the above findings, we have
ignored the coupling to the magnetic vector potential, so that Eq.\
(\ref{effprime}) essentially describes a superfluid.  Although the
complete effective theory is Galilei invariant \cite{GWW,nr}, the
linearized form (\ref{effprime}) is invariant under Lorentz
transformations, with $c$ replacing the speed of light.

In the dual formulation, where the roles of charges and vortices are
interchanged, the Anderson-Bogoliubov mode mediating the interaction
between two vortices is represented as a photon associated with a
fictitious gauge field $a_\mu$, i.e., (in relativistic notation)
$\partial_\mu \varphi \sim \epsilon_{\mu \nu \lambda} \partial^\nu
a^\lambda$.  In 2+1 dimensions, a photon has only one transverse
direction and thus only one degree of freedom---as has the
Anderson-Bogoliubov mode.  The elementary excitations of the dual theory
are the vortices, described by a complex scalar field $\psi$.
Specifically, the (well-known) dual theory of Eq.\ (\ref{effprime}) is
the Ginzburg-Landau model \cite{dualGL,GFCM,WeZe,KKS}
\begin{equation} 
{\cal L}_{\rm dual} = -\case{1}{4} f_{\mu \nu}^2 + |(\partial_\mu -i g
a_\mu) \psi|^2 - m_\psi^2 |\psi|^2 - \case{1}{4} u|\psi|^4,
\end{equation} 
with $f_{\mu \nu} = \partial_\mu a_\nu - \partial_\nu a_\mu$, $m_\psi$ a
mass parameter, and $u$ the strength of the self-coupling.  Both the
gauge part as well as the matter part of the dual theory are of a
relativistic form.  The gauge part is because the effective theory
(\ref{effprime}) is Lorentz invariant, while the matter part is because
vortices of positive and negative circulation can annihilate, and can
also be created.  In this sense they behave as relativistic particles.
As was pointed out in Ref.\ \cite{WeZe}, the speed of ``light'' in the
gauge and matter part in general differ.

The interaction potential (\ref{V2d}) between two external vortices is
now being interpreted as the 2-dimensional Coulomb potential between
charges.  The observation concerning the critical behavior of the
Ginzburg-Landau model implies that the qualitative change in $V(r)$
upon entering the critical region is properly represented in the dual
formulation.

Whereas in the conducting phase, the charges are condensed, in the
insulating phase, the vortices are condensed \cite{MPAFisher}.  In the
dual theory, the vortex condensate is represented by a nonzero
expectation value of the $\psi$ field, which in turn leads via the
Anderson-Higgs mechanism to a mass term for the gauge field $a_\mu$.
Because $(\epsilon_{\mu \nu \lambda} \partial^\nu a^\lambda)^2 \sim
(\partial_\mu \varphi)^2$, the mass term $a_\mu^2$ with two derivatives
less implies that the Anderson-Bogoliubov mode has acquired an energy
gap.  That is to say, the phase where the vortices are condensed is
indeed an insulator.  Since electric charges are seen by the dual theory
as flux quanta, they are expelled from the system as long as the dual
theory is in the Meissner state.  Above the critical field $h =
\nabla_\perp \times {\bf a} = h_{c_1}$ they start penetrating the system
and form an Abrikosov lattice.  In the original formulation, this
corresponds to a Wigner crystal of the charges.  Finally, when more
charges are added and the dual field reaches the critical value
$h_{c_2}$, the lattice melts and the charges condense in the superfluid
phase described by the effective theory (\ref{effprime}).

\acknowledgements I'm grateful to M. Krusius for the kind hospitality at
the Low Temperature Laboratory in Helsinki.  I wish to acknowledge
useful conversations with J.\ Hove, N.\ Kopnin, M.\ Krusius, K.\ Nguyen,
M.\ Paalanen, A.\ Sudb\o, and especially G. Volovik.

This work was funded in part by the EU sponsored programme Transfer and
Mobility of Researchers under contract No.\ ERBFMGECT980122.

\end{multicols}

\begin{thebibliography}{99}
\bibitem{note} Note that only when the anomalous scaling dimension
$d_{\rho_{\rm s}} = (d+z)-2$ vanishes, the form of the interaction
between two vortices remains unchanged upon entering the critical
region.  An example is given by the Berezinskii-Kosterlitz-Thouless
phase transition where $d=2,z=0$.
\end{thebibliography}
\end{document}